\documentclass[12pt,draftclsnofoot,onecolumn]{IEEEtran}
\usepackage{amssymb}
\usepackage{amsfonts}
\usepackage{amsfonts,subfigure,multicol,color,verbatim,
graphicx,cite,epsfig,amssymb,amsmath,cases,bm,algorithm,
algorithmic,xcolor,multirow,array}
\usepackage{multirow}
\usepackage{multicol}
\usepackage{supertabular}
\usepackage{array}
\usepackage{colortbl}
\usepackage{longtable}

\begin{document}
\markboth{Submit to IEEE Network, vol. x, no. y, Mon. 2015} {Peng:
C-RANs \ldots}

\title{Fog Computing based Radio Access
Networks: Issues and Challenges}

\author{
Mugen~Peng, \IEEEmembership{Senior Member,~IEEE}, Shi~Yan,
Kecheng~Zhang, and Chonggang~Wang,~\IEEEmembership{Senior
Member,~IEEE}
\thanks{Mugen~Peng (e-mail: {\tt pmg@bupt.edu.cn}), Shi~Yan (e-mail: {\tt yanshi01@bupt.edu.cn}), and Kecheng~Zhang (e-mail: {\tt pdjzzkc@gmail.com}) are with the Key Laboratory of Universal Wireless Communications for
Ministry of Education, Beijing University of Posts and
Telecommunications, Beijing, China. Chonggang~Wang (e-mail: {\tt
cgwang@ieee.org}) is with the InterDigital Communications, King of
Prussia, PA, USA.}
\thanks{This work was supported in part by the National Natural Science Foundation of China (Grant No. 61222103, No.61361166005), the National Basic Research Program of China (973 Program) (Grant No. 2013CB336600), and the Beijing Natural Science Foundation (Grant No. 4131003).}
\thanks{The final paper was submitted on \today.}}

\date{\today}
\renewcommand{\baselinestretch}{1.5}
\thispagestyle{empty} \maketitle \thispagestyle{empty}
\vspace{-15mm}
\begin{abstract}
A fog computing based radio access network (F-RAN) is presented in
this article as a promising paradigm for the fifth generation (5G)
wireless communication system to provide high spectral and energy
efficiency. The core idea is to take full advantages of local radio
signal processing, cooperative radio resource management, and
distributed storing capabilities in edge devices, which can decrease
the heavy burden on fronthaul and avoid large-scale radio signal
processing in the centralized baseband unit pool. This article
comprehensively presents the system architecture and key techniques
of F-RANs. In particular, key techniques and their corresponding
solutions, including transmission mode selection and interference
suppression, are discussed. Open issues in terms of edge caching,
software-defined networking, and network function virtualization,
are also identified.
\end{abstract}


\newpage

\section{Introduction}

Compared to the fourth generation (4G) wireless communication
system, the fifth generation (5G) wireless communication system
should achieve system capacity growth by a factor of at least 1000,
and the energy efficiency (EE) growth by a factor of at least 10
\cite{I:DP}. To achieve these goals, the cloud radio access network
(C-RAN) has been proposed as a combination of emerging technologies
from both the wireless and the information technology industries by
incorporating cloud computing into radio access networks (RANs)
\cite{I:5G}. C-RANs have come with their own challenges in the
constrained fronthaul and centralized baseband unit (BBU) pool. A
prerequisite requirement for the centralized processing in the BBU
pool is an inter-connection fronthaul with high bandwidth and low
latency. Unfortunately, the practical fronthaul is often capacity
and time-delay constrained, which has a significant decrease on
spectral efficiency (SE) and EE gains.

To overcome the disadvantages of C-RANs with the fronthaul
constraints, heterogeneous cloud radio access networks (H-CRANs)
have been proposed in \cite{I:Peng_HCRAN}. The user and control
planes are decoupled in such networks, where high power nodes (HPNs)
are mainly used to provide seamless coverage and execute the
functions of control plane, while remote radio heads (RRHs) are
deployed to provide high speed data rate for the packet traffic
transmission in the user plane. HPNs are connected to the BBU pool
via the backhaul links for interference coordination. Unfortunately,
H-CRANs are still challenging in practice. First, since the location
based social applications become more and more popular, the traffic
data over the fronthaul between RRHs and the centralized BBU pool
surges a lot of redundant information, which worsens the fronthaul
constraints. Besides, H-CRANs do not take full advantage of
processing and storage capabilities in edge devices, such as RRHs
and ``smart'' user equipments (UEs), which is a promising approach
to successfully alleviate the burden of the fronthaul and BBU pool.
Moreover, operators need to deploy a huge number of fixed RRHs and
HPNs in H-CRANs to meet requirements of peak capacity, which makes a
serious waste when the volume of delivery traffic is not
sufficiently large. To solve such challenges, revolutionary
approaches involving new radio access network architectures and
advanced technologies need to be explored.

Fog computing is a term for an alternative to cloud computing that
puts a substantial amount of storage, communication, control,
configuration, measurement and management at the edge of a network,
rather than establishing channels for the centralized cloud storage
and utilization, which extends the traditional cloud computing
paradigm to the network edge \cite{I:fogcomputing}. Note that it is also called edge cloud computing, which pushes applications, data and computing content away from centralized points to the logical extremes of a network. In this article, we unify them as the fog computing. Based on fog
computing, the collaboration radio signal processing (CRSP) can not
only be executed in a centralized BBU pool in H-CRANs, but also can
be hosted at RRHs and even wearable ``smart'' UEs. To efficiently
support and integrate new types of ``smart'' UEs, the on-device
processing and cooperative radio resource management (CRRM) with a
little distributed storing should be exploited. Meanwhile, from the
viewpoint of mobile applications, UEs do not have to connect to the
BBU pool to download the packet if the applications happen locally
or the same content has stored in adjacent RRHs. Inspired by these
characteristics of fog computing, to alleviate the existing
challenges of H-CRANs, and take full advantages of local caching,
CRSP and CRRM functions at edge devices, including RRHs and ``smart"
UEs, the fog computing based RAN (F-RAN) architecture is proposed in
this article.

There are some apparent advantages in F-RANs, including the
real-time CRSP and flexible CRRM at the edge devices, the rapid and
affordable scaling that make F-RANs adaptive to the dynamic traffic
and radio environment, and low burden on the fronthaul and BBU pool.
Furthermore, the user-centric objectives can be achieved through the
adaptive technique among device to device (D2D), wireless relay,
distributed coordination, and large-scale centralized cooperation.
To incorporate fog computing in edge devices, the traditional RRH is
evolved to the fog computing based access point (F-AP) through equipping
with a certain caching, CRSP, and CRRM capabilities.

Much attention has been paid to the mobile fog computing and edge
cloud from the viewpoint of information sciences and Internet of
Things (IoT) recently. In \cite{I:fogcomputing}, a fog computing
platform to deliver a rich portfolio of new services and
applications at the network edge has been proposed by Cisco firstly.
The design of mobile fog as a programming model for large-scale,
latency-sensitive applications in the IoT has been introduced in
\cite{I:fogcomputing2}. Lately, in \cite{I:edgecloud}, an edge cloud
and underlay network architecture has been proposed and its
corresponding simulation performances for the edge caching has been
presented. To the best of our knowledge, there is still no published
work to discuss the fog computing in 5G RANs.

In this article, we are motivated to make an effort to offer a
comprehensive discussion on system architectures and technological
principles in F-RANs. Specifically, the F-RAN system architecture is
presented, where the new communication entity F-AP is defined and
the software-defined F-RAN architecture is designed. An adaptive
transmission mode selection and interference suppression in F-RANs
are researched, and the corresponding performances are discussed.
The future challenges and open issues are presented as well.

The remainder of this paper is outlined as follows. F-RAN system
architectures will be introduced in Section II. The adaptive
transmission mode selection will be presented in Section III. The
interference suppression techniques will be discussed in Section IV.
Future challenges and open issues will be highlighted in Section V,
followed by conclusion in Section VI.

\section{F-RAN System Architecture}

\begin{figure}[!htp]
\centering
\includegraphics[scale=0.20]{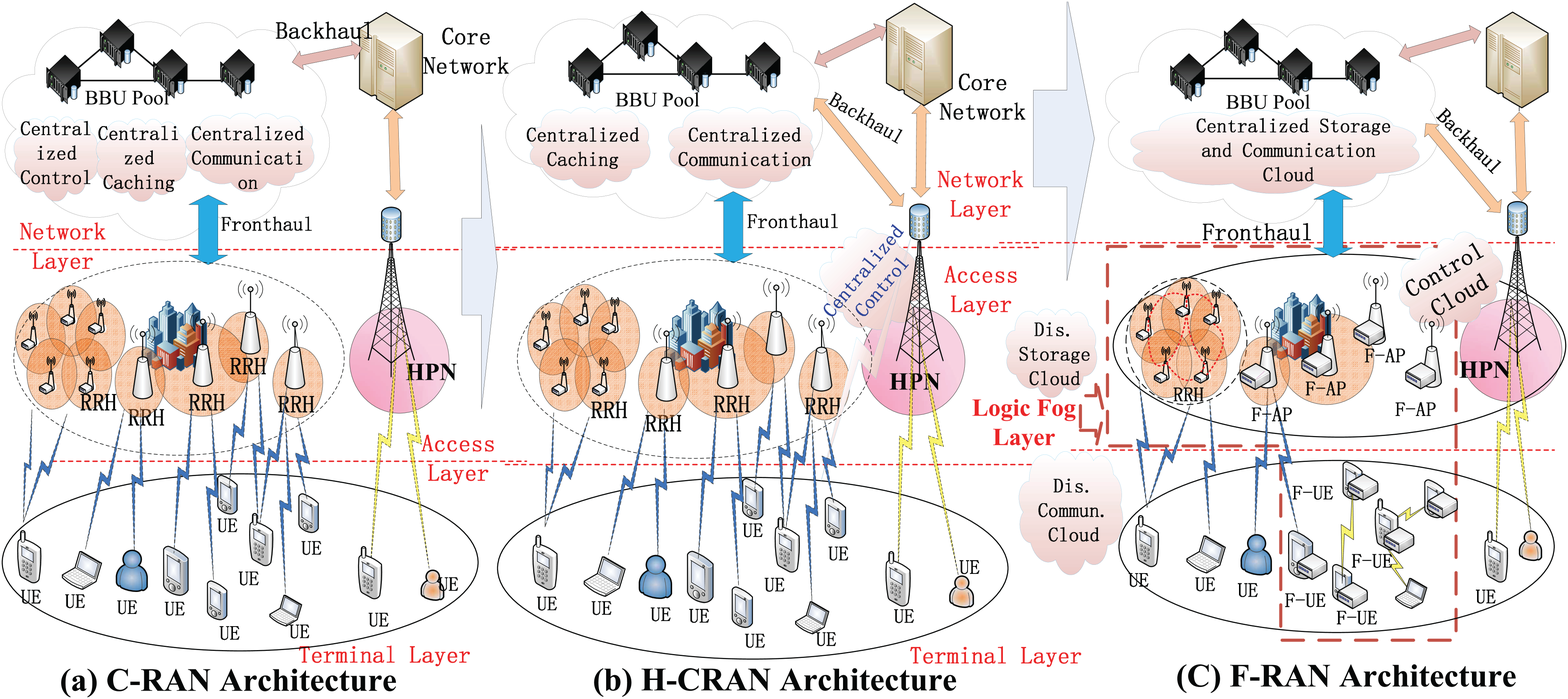}
\caption{System architecture evolution through F-RANs}\label{fig_FRAN_evolution}\vspace*{-1.5em}
\end{figure}

The F-RAN system architecture evolution from C-RAN is proposed in
Fig. \ref{fig_FRAN_evolution}. In C-RAN and H-CRAN system architectures, all CRSP functions and the
application storing are centralized at the cloud server, which
require billions of UEs transmit and exchange their data fast enough
with the BBU pool. Note that the main difference between C-RANs and H-CRANs is that the centralized control function is shifted from the BBU pool in C-RANs to the HPN in H-CRANs. Meanwhile, some UEs can access the HPN to alleviate the burdens on the fronthaul of C-RANs. Furthermore, two biggest problems in both C-RANs and H-CRANs are the big transmission
delay and heavy burden on the fronthaul. A simple solution is to
stop transmitting all the torrent of data to the BBU pool and
process part of the radio signals at the local RRHs and even
``smart" UEs. Meanwhile, to avoid all traffics are offloaded
directly from the centralized cloud server, some local traffics
should be delivered from the caching of adjacent RRHs (denoted by
F-APs in F-RANs) or ``smart" UEs (denoted by F-UEs) to save the
spectral usage of constrained fronthaul and decrease the
transmission delay.

As shown in Fig. \ref{fig_FRAN_evolution} (c), some distributed communication and storage functions exist in the logic fog layer.
Accurately, the proposed F-RAN takes full advantages of the convergence of
cloud computing, heterogeneous networking, and fog computing. Four kinds of clouds are defined: global
centralized communication and storage cloud, centralized control
cloud, distributed logical communication cloud, and distributed
logical storage cloud. The global centralized communication and
storage cloud is same as the centralized cloud in C-RANs, and the
centralized control cloud is used to complete functions of control
plane and located in HPNs. The distributed logical communication
cloud located in F-APs and F-UEs are responsible for the local CRSP
and CRRM functions, while the distributed logical storage cloud
represents the local storing and caching in edge devices.

\begin{figure}[!htp]
\centering
\includegraphics[width=4.5in]{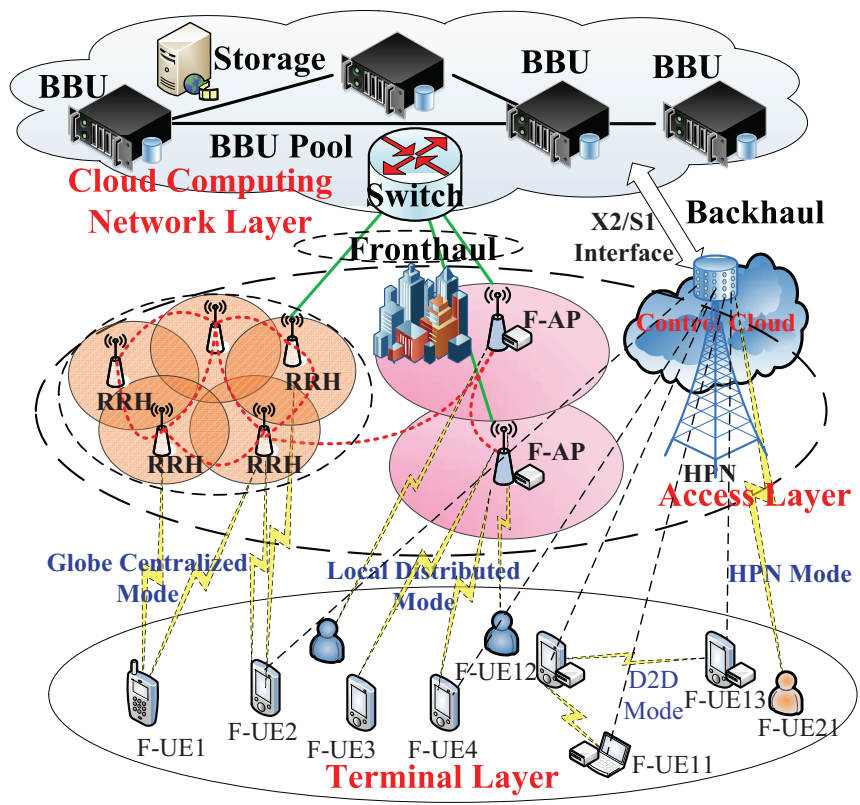}
\caption{System model for implementing F-RANs}\label{fig_FRAN_model}\vspace*{-1.5em}
\end{figure}


The proposed system model for implementing F-RANs illustrated in
Fig. \ref{fig_FRAN_model} can be regarded as the practical example that
implement four kinds of clouds defined in Fig. \ref{fig_FRAN_evolution} (c), which
comprises of terminal layer, network access layer, and cloud
computing layer. Accurately, the F-APs and F-UEs in the terminal
layer and network access layer formulate the fog computing layer. In
the terminal layer, adjacent F-UEs can communicate with each other
through the D2D mode or the F-UE based relay mode. For example, F-UE13
and F-UE11 can communicate with each other with the help of F-UE12, in
which F-UE12 can be regarded as a mobile relay. If there are some data
to be directly transmitted between F-UE11 and F-UE12, the D2D mode is
used. The network access layer consists of F-APs and HPNs.
Similarly with H-CRANs, all F-UEs access the HPN to obtain all
system information related signalling, which fulfills the functions
of the control plane. In addition, F-APs are used to forward and
process the received data. F-APs are interfaced to the BBU pool in
the cloud computing layer through the fronthaul links, while HPNs
are interfaced to the BBU pool with the backhaul links, which
indicates that the signals over fronthaul links should be
large-scale processed in the BBU pool, and there are only
coordinations between the BBU pool and HPN over the backhaul links.
The traditional X2/S1 interface for the backhaul link is backward
compatible with that defined in 3$^{rd}$ generation partnership
project (3GPP) standards for long term evolution (LTE) and
LTE-Advanced systems. The BBU pool in the cloud computing layer is
compatible with that in H-CRANs. Meanwhile, the centralized caching
is located in the cloud computing layer. Since a large number of
CRSP and CRRM functions are shifted to F-APs and F-UEs, the burden
on the fronthaul and BBU pool are alleviated. Furthermore, the
limited caching in F-APs and F-UEs can make some packet service
offload from edge devices, and not from the centralized caching. The
characteristics of F-RANs compared with C-RANs and H-CRANs are
illustrated in Tab. \ref{Table1}.

\begin{small}
\begin{longtable}[c]{|m{.2\linewidth}|m{.2\linewidth}|m{.25\linewidth}|m{.25\linewidth}|}
\caption{Advantage Comparisons of C-RANs, H-CRANs, and F-RANs}\label{Table1}\\
  \hline
  Items & C-RANs & H-CRANs & F-RANs\\
  \hline
  Burden on Fronthaul and BBU pool & Heavy & Medium & Low \\
  \hline
  Latency & High & High & Low \\
  \hline
  Decouple of user and control planes & No & Yes & Yes \\
  \hline
  Caching and CRSP & Centralization & Centralization & Mixed Centralization and Distribution \\
  \hline
  CRRM & Centralization & Centralization, and Distribution between the BBU pool and HPNs & Mixed Centralization and Distribution \\
  \hline
  Performance Gains & Fronthual constraint & Fronthual and backhaul constraint & Backhaul constraint \\
  \hline
  Implementing Complexity & High in the BBU pool, low in RRHs and UEs & High in the BBU pool, low in RRHs and UEs & Medium in the BBU pool, F-APs, and F-UEs\\
  \hline
  Traffic Characteristic & Packet service & Packet service, real-time voice service & Packet service, real-time voice service \\
  \hline
\end{longtable}
\end{small}

\subsection{Key Components: F-APs and F-UEs}

To execute the necessary CRSP and CRRM functions in the logic
fog-computing layer adaptively, or offload the delivered packet
traffic from edge devices locally, the evolved F-APs and F-UEs are
presented to enhance the traditional RRHs and UEs, respecitvely.
F-APs are mainly used to process local CRSP and CRRM for accessed
F-UEs, offer interference suppression and spectral sharing for D2D
F-UEs, and compress-and-forward the received information to the BBU
pool through fronthaul. F-AP integrates not only the front radio
frequency (RF), but also the local distributed CRSP and simple CRRM
functions. By processing collaboratively among multiple adjacent
F-APs, the overload of the fronthaul links can be released, and the
queuing and transmitting latency can be alleviated. When all CRSP
and CRRM functions are shifted to the BBU pool, F-AP is degenerated
to a traditional RRH. F-UEs denote UEs accessing F-APs and working
in the D2D mode.

Since the proposed F-RAN is evolved from HetNets and C-RANs, it is fully compatible with the other 5G systems. The advanced 5G techniques, such as the massive multiple-input multiple-output (MIMO), cognitive radio, millimeter wave communications, and non-orthogonal multiple access, can be used directly in F-RANs. The local distributed CRSP techniques among adjacent F-APs
inherited from the virtual MIMO can achieve high diversity and multiplexing gains for F-UEs without consuming fronthaul links. Herein, the interference among adjacent
F-APs can be suppressed in a distributed manner under the help of
coordinated multiple points (CoMP) transmission and reception. If
the interference is not tackled efficiently by the local distributed
CRSP and CRRM techniques, the global centralized CRSP and CRRM are
triggered with the assistance of the BBU pool, which is the same way
in C-RANs. When the traffic load is low, some potential F-APs fall
into the sleep mode. While the traffic load becomes tremendous in a
small special zone, F-APs and HPNs are active to absorb the high
capacity, and the D2D or F-UE based relay modes can be further
triggered to meet the huge capacity requirement.

\subsection{Hierarchical Architectures}

The hierarchical architecture of F-RANs consisting of fog computing
layer and cloud computing layer can make F-UEs adaptively work in
the optimal mode. In the terminal layer, an F-UE can directly
communicate with the adjacent F-UE in the D2D mode without the
assistance of F-APs, in which the HPN is used to deliver overall
control signalling for the D2D paired F-UEs. By reusing the same
radio resources with F-UEs connecting with F-APs, the D2D mode is
particularly beneficial to satisfy the demand of high data-rate
transmission, and also capable of enhancing the overall throughput.
However, the D2D mode is severely constrained by the communication
distance and the capability of F-UEs, in which the traditional UEs
without supporting D2D mode can not be supplied. If the
communication distance of two potential paired F-UEs is beyond the
D2D distance threshold, the F-UE based relay mode will be triggered
to provide the communication for these two F-UEs with the third F-UE
close to them.

\begin{figure}[H]
\centering \makeatletter\def\@captype{figure}\makeatother
\subfigure[The mesh topology for the F-AP
connecting]{\label{FRRH_forming:a}
\includegraphics[width=2.6in]{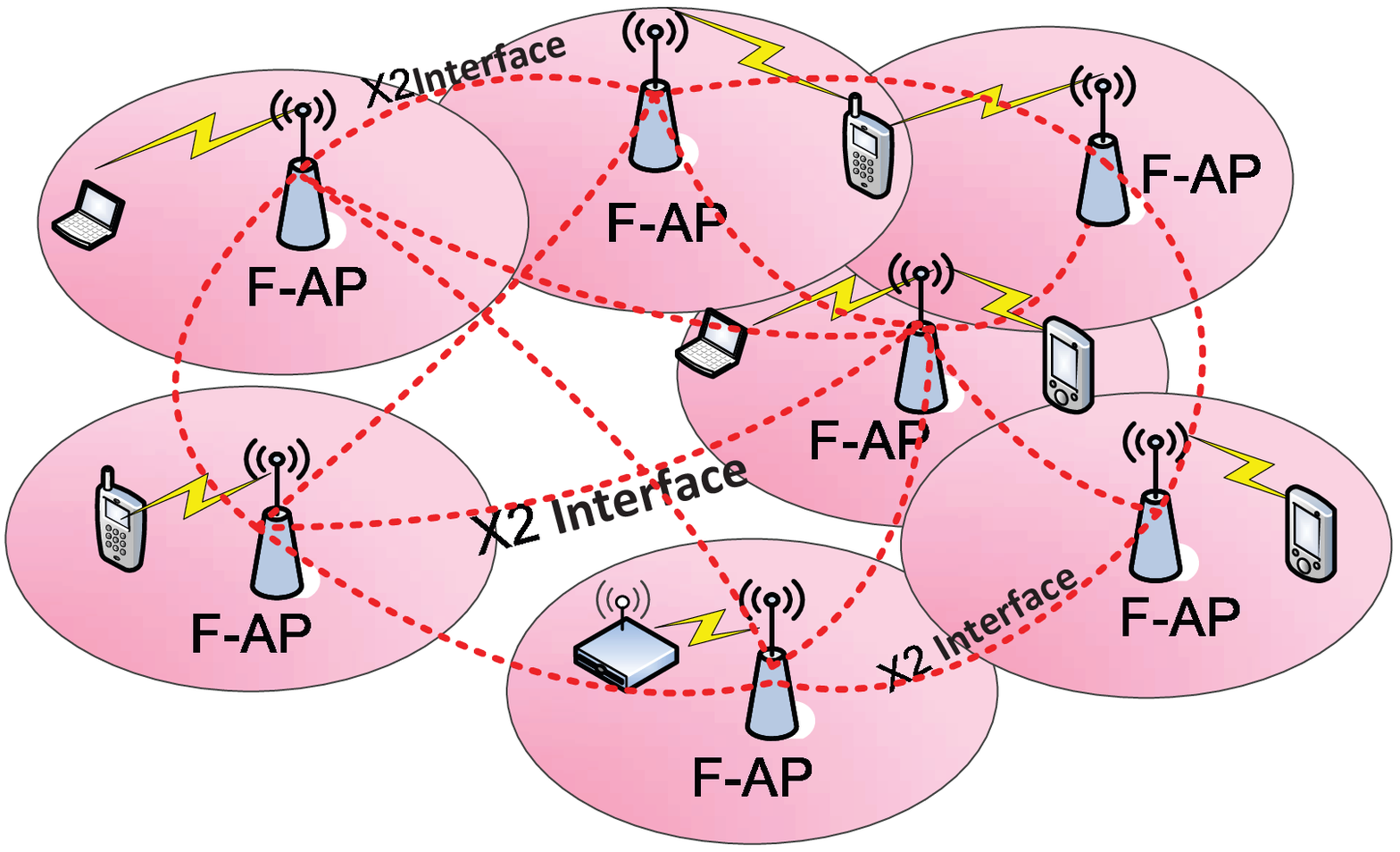}} \hspace{0.1 in}
\subfigure[The tree-like topology for the F-AP connecting]{
\label{FRRH_forming:b}
\includegraphics[width=2.6in]{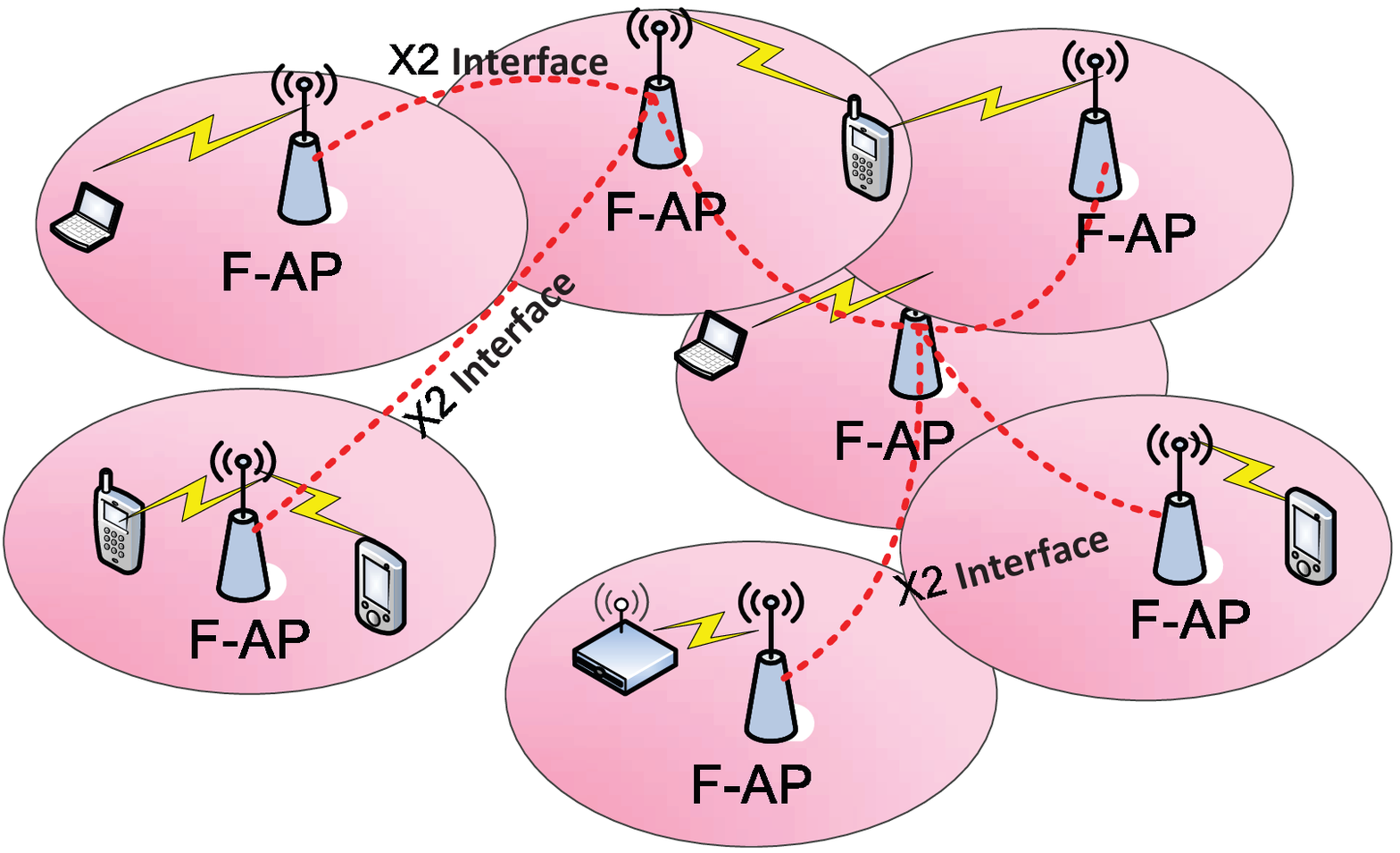}}
\caption{Two kinds of F-AP connecting topology in the network
access layer} \label{FRRH_forming}
\end{figure}

In the network access layer, there are two types of edge
communication entities: HPN and F-AP. Inherited from H-CRANs, the
HPN is mainly used to deliver the overall control signalling and
provide seamless coverage with basic bit rate for high mobile F-UEs.
HPNs with massive MIMO are still critical to guarantee the backward
compatibility with the existing wireless systems. The overall
control channel overhead and cell specific reference signals for
F-RANs are delivered by HPNs, and thus F-RANs can decrease the
unnecessary handover and alleviate the synchronous constrains. If
the CRSP and CRRM functions are ended in F-APs, they have the same
functions of small cell base stations, in which the distributed
interference coordination like the CoMP is adopted to suppress the
intra-tier and inter-tier interference. In addition, the adjacent
F-APs are interconnected and formed into different kinds of
topology to implement the local distributed CRSP, as shown in Fig.
\ref{FRRH_forming}, in which each F-AP is connected to the others
with the data and control interfaces S1 and X2, respectively. The
interference among adjacent F-APs can be suppressed by the
distributed CRSP and CRRM without the assistance of BBU pool. In
Fig. \ref{FRRH_forming:a}, the adjacent F-APs are formed into a
mesh-topology group to implement the distributed CRSP and deliver
the packet traffic stored herein. In Fig. \ref{FRRH_forming:b}, the
tree-like topology for the F-AP connecting is illustrated. Both
mesh and tree-like topologies can decrease negative influences of
capacity-constrained fronthaul links. Compared with the mesh
topology, the wireless cluster feasibility in the tree-like topology
is about 50 percent lower with significant reduced network
deployment and maintenance cost \cite{ConnectedRRH}. Therefore, the
tree-like topology is preferred in the practical F-RANs.

To achieve large-scale centralized CRSP and CRRM gains, F-APs are simplified into the traditional RRHs, which forward received signals from UEs to the the BBU pool.
It is noted that the fronthaul constraints challenging to C-RANs and
H-CRANs are significantly alleviated in F-RANs because many CRSP and
CRRM functions are executed in F-APs and F-UEs. Furthermore, many
delivered packet traffic are stored not in the cloud server but in
the edge devices.

The cloud computing layer is software defined and is characterized
by attributes of centralized computing and caching, which is
inherited from C-RANs. All signal-processing units work together in
a large physical BBU pool to share the overall F-RAN's signaling,
traffic data as well as channel state information. When the network
load grows, the operator only needs to upgrade the BBU pool to
accommodate the increased capacity. The BBU pool is much easier to
implement the joint processing and scheduling to coordinate the
inter-tier interferences between F-APs/RRHs and HPNs via the centralized
large-scale CoMP approach. The centralized caching is used to store
all global packet traffic.

\section{Transmission Mode Selection}

\begin{figure}[!htp]
\centering
\includegraphics[width=4.5in]{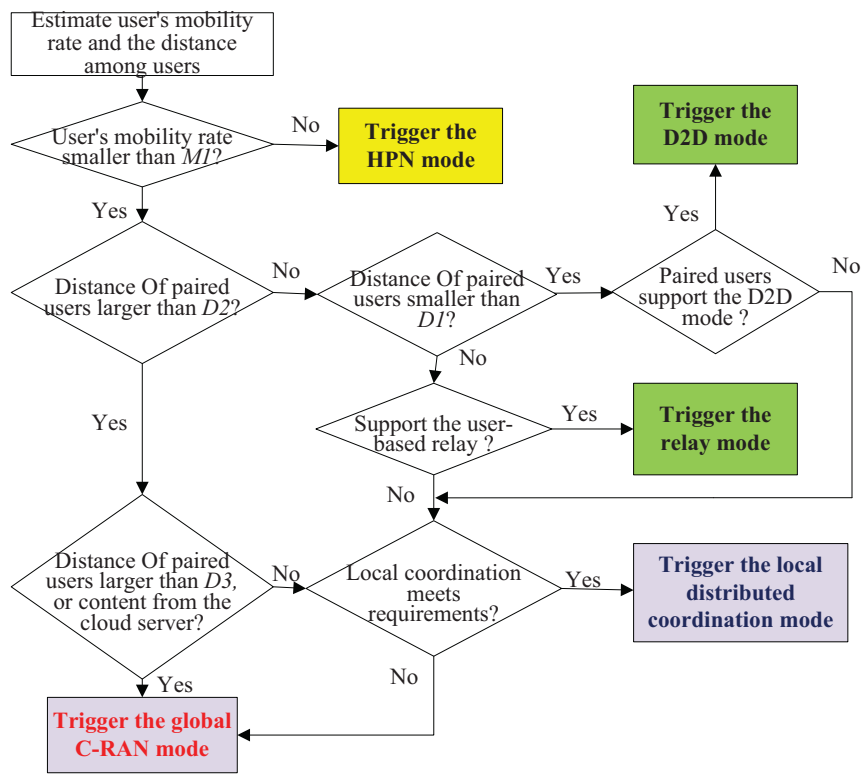}
\caption{Adaptive transmission mode selection in
F-RANs}\label{fig_Modeselection}\vspace*{-1.5em}
\end{figure}

UEs access to the F-RAN adaptively, and there are four transmission
modes to be selected according to F-UEs' movement speed,
communication distance, location, quality of service (QoS)
requirements, processing and caching capabilities: D2D and relay
mode, local distributed coordination mode, global C-RAN mode, and
HPN mode. In the D2D and relay mode, two F-UEs communicate with each
other via the D2D or the UE-based wireless relay techniques. The
local distributed coordination mode means that F-UEs access to the
adjacent F-AP and the communication is ended herein. The global
C-RAN mode means that all CRSP and CRRM functions are implemented
centrally at the BBU pool, which is same as that done in C-RANs.
Similarly with H-CRANs, F-UEs with high movement speed or in the
coverage hole of F-APs have to access the HPN, which is denoted by
the HPN mode. As illustrated in Fig. \ref{fig_Modeselection}, an
adaptive mode selection is presented in this article take full
advantages of these four modes.

Except that all F-UEs periodically listen to the delivered control
signalling, the optimum transmission mode is selected by the desired
F-UE under the supervising of HPNs. To determine the optimal
transmission mode for each F-UE, the movement speed of F-UEs and the
distance of different F-UE pairs are estimated according to the
public broadcast pilot channels from HPNs firstly. If a F-UE is in a
high-speed mobile state or provides the real-time voice
communication, the HPN mode is triggered with a high priority. If
two F-UEs communicating with each other have a slow relative
movement speed and their distance is not bigger than the threshold
$D1$, the D2D mode is triggered. Otherwise, if their distance is
bigger than $D1$ but smaller than $D2$, and there is one adjacent
F-UE friendly acting as the F-UE based relay for these two UEs to
achieve better performances than the other modes, the F-UE based
relay mode is triggered. Furthermore, if two desired F-UEs move
slowly, and 1) their distance is larger than $D2$ but smaller than
$D3$, or 2) their distance is not bigger than $D2$, but at least one
F-UE does not support the D2D and relay mode, the local distributed
coordination mode is adopted. If the local distributed coordination
mode can not afford the expected performance, or the distance
between two desired F-UEs is bigger than $D3$, or the delivered
content comes from the cloud server, the global C-RAN mode is
triggered.

\subsection{D2D and Relay Mode}

\begin{figure}
\centering  \vspace*{0pt}
\includegraphics[scale=0.7]{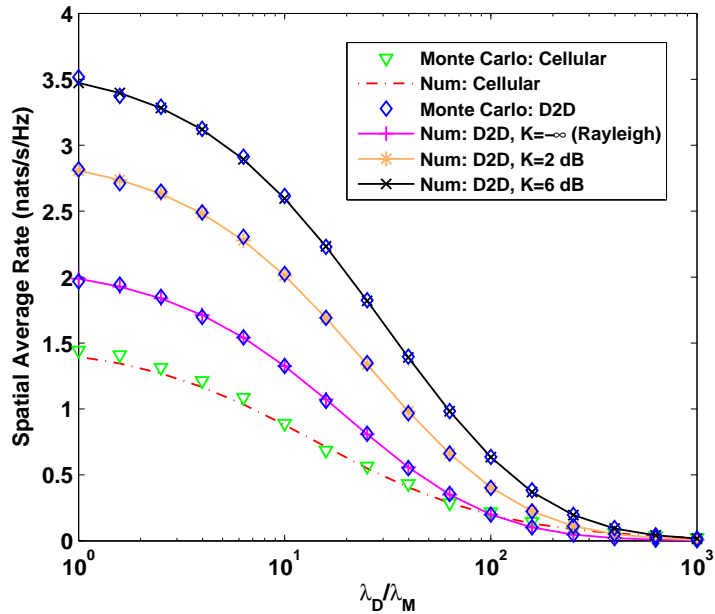}
\setlength{\belowcaptionskip}{-100pt} \caption{Spatial average rate
for the cellular and D2D transmissions.}
\label{Ergodic_Capacity_Fig_5}\vspace*{-10pt}
\end{figure}

The D2D and relay mode is a user-centric access strategy which makes
the communication only exist in the terminal layer, which can
achieve significant performance gains benefitting from the D2D or
wireless relay techniques, and effectively relieve the burden on the
fronthaul. In this mode, the HPN assigns the device identification
for each F-UE. With necessarily lower antenna heights in D2D
communication links, the fast fading channels are likely to contain
strong line-of-sight components, which are different from the
Rayleigh fading distribution in conventional wireless networks. In
\cite{D2D_peng}, the performance gains of D2D over cellular
transmissions by taking into account the fading channel propagations
with different Rician $K$-factors are analyzed and evaluated. As
shown in Fig.~\ref{Ergodic_Capacity_Fig_5}, the numerical
performance results of spatial average rate for both the cellular
transmission and the D2D transmission are validated. It illustrates
how the spatial average rate varies with the increase of the density
of D2D users ${\lambda _{\mathrm{D}}}$ when the density of HPNs
${\lambda _{\mathrm{M}}}$ is fixed. In comparison with the Rayleigh
fading channels, nearly 38\% performance gains can be achieved when
$K$ = 2 dB in the low ${\lambda _{\mathrm{D}}}$ region, while nearly
68\% performance gains are observed in the case of $K$ = 6 dB.
However, the spatial average rate performance is severely degraded
for any scenario when the density of ${\lambda _{\mathrm{D}}}$ is
sufficiently large. The performance gains from the F-UE based relay
mode can be referred to the two-way relay with network coding
approaches in \cite{Networkcoding_peng}.

\subsection{Local Distributed Coordination Mode}

To decrease the burden on fronthaul and suppress the interference
timely, or directly offload the traffic from not the cloud server
but F-APs, the local distributed coordination mode is used and the
corresponding performance gains of interference coordination mainly
source from CoMP. The F-AP cluster to execute the local distributed
coordination mode is adaptively formed, which takes the implementing
complexity and CoMP gains into account. The CoMP gains strictly
depend on the topology of F-RAN cluster and the backhaul capacity of
connecting F-APs. Based on the local distributed coordination mode,
a remarkable increase of SE is achievable especially for the
cell-edge user performance, which is on the order of 70\% for the
downlink and 122\% in the uplink \cite{I:edgecloud}. Meanwhile, the
F-AP association is critical to improve SE. F-UE tries to access
the optimal F-AP with the maximal received power strength when its
minimum QoS can be satisfied. If the optimal F-AP has no sufficient
radio resource that can be allocated to the F-UE or the interference
to the adjacent F-APs is bigger than the predefined threshold, the
F-UE tries to access the sub-optimal F-AP with allowances. It is
noted that only single F-AP is allowed to access for each F-UE in
this mode.

\subsection{Global C-RAN Mode}

If F-APs that should be coordinated to suppress the interference
for the desired UE are not inter-connected, or the content delivered
to the desired UE is only stored in the cloud server, the global
C-RAN mode is adopted. In this mode, RRHs forward the received
radio signals to the BBU pool, and the BBU pool executes all CRSP
and CRRM functions globally and centrally, which is the same way as
that in C-RANs. Different from the local distributed coordination
mode, several RRHs can work together for the desired UE to
improve SE. Under the help of other three modes, the capacity
demands on the fronthaul have significant decreases, which
alleviates the capacity and latency constraints.

\subsection{HPN Mode}

If the content delivered to the desired UE is bursty with a low
volume, or the movement speed is beyond the pre-defined threshold,
the HPN mode is preferred, which can decrease the overhead of
control channel, and avoid the unnecessary handover. This mode is
mainly used to provide the seamless coverage with the basic QoS
support. The resource sharing and performance gains of HPN mode are
presented in \cite{HCRAN_TVT_peng}, where an enhanced soft
fractional frequency reuse (S-FFR) scheme can be used to mitigate
the inter-tier interference between HPNs and F-APs. Only partial
radio resources are allocated to UEs with low QoS requirements,
and the remaining radio resources are allocated to F-UEs accessing
F-APs with high QoS requirements. UEs with low QoS requirements
accessing HPNs share the same radio resources with F-UEs accessing
F-APs. The analysis and simulation results show that this S-FFR
scheme has a significant SE and EE gains in \cite{HCRAN_TVT_peng}.

\section{Interference Suppression}

Since the same radio resources are shared among F-UEs with these
four transmission modes, the severe interference is challenging to
improve performances of F-RANs. Inspired by the CoMP in 3GPP, the
interference suppression techniques in F-RANs can be categorized
into coordinated precoding and coordinated scheduling. The
coordinated precoding technique is utilized to decrease interference
with centralized and distributed manners in the physical layer for
the global C-RAN mode and the local distributed coordination mode,
respectively. The coordinated scheduling is mainly used to suppress
the interference in the medium access control (MAC) layer.


\subsection{Coordinated Precoding}

\begin{figure}[t!]
\centering
\includegraphics[width=5.0in]{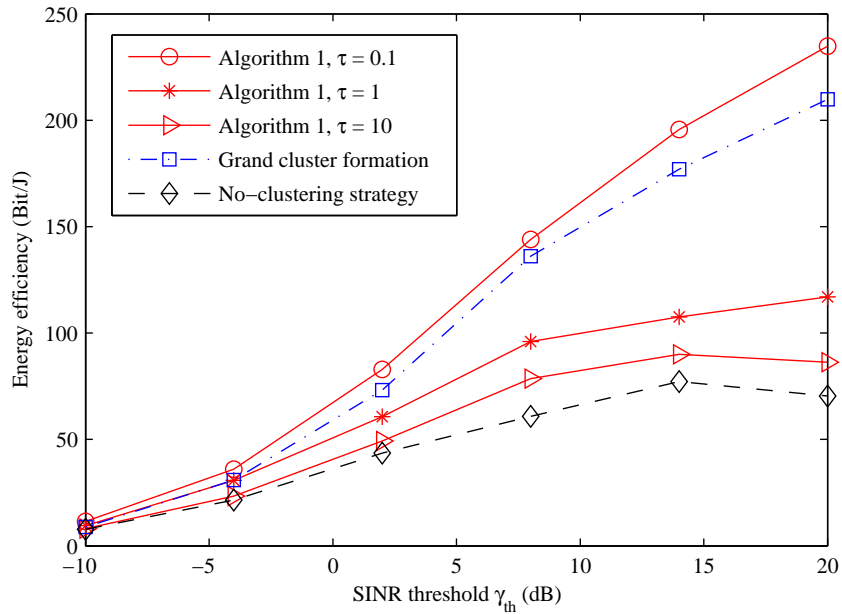}
\caption{Energy efficiency of Algorithm 1. The energy exponent in
the utility function is set as $\tau = 0.1,~1,$ and 10, and the
density of F-APs is $\lambda = 10^{-5}$.} \label{fig:ee}
\end{figure}

The coordinated precoding technique is generally categorized into
the global and distributed manners. The global coordinated precoding
include the massive MIMO for the HPN mode with a large-scale
antennas at a single site, and the large-scale cooperative MIMO for
the global C-RAN mode with distributed F-APs located in different
sites. The distributed coordinated precoding technique refers to the
joint processing CoMP with distributed F-APs in the same cluster
for the local distributed coordination mode. To balance performances
and complexity, the coordinated precoding size should be sparsely
designed, in which only a small fraction of the overall entries in
the channel matrix have reasonably large gains, and they can be
ignored leading to a great reduction in processing complexity and
channel estimation overhead. In \cite{Zhaozhongyuan_clustsize}, the
coordinated precoding cluster formation for F-APs is studied, and
an explicit expression of the successful access probability for a
fixed intro-cluster cooperation strategy is derived by applying
stochastic geometry. By using the obtained theoretical result as a
utility function, the problem of grouping F-APs is formulated as a
coalitional formation game, and then the intro-cluster cooperation
algorithm (called Algorithm 1 herein) based on the merge and split
approaches is obtained. To estimate the performance gains, the grand
cluster formation and the no-clustering strategies, which can show
the performance of the completely centralized and the completely
distributed schemes, respectively, are chosen as two baseline
schemes. As shown in Fig. \ref{fig:ee}, the impact of the power
consumption part is mitigated when $\tau = 0.1$, which can provide
flexible choices for the cluster size settings. In this
circumstance, the target data rate increases as the signal to
interference plus noise ratio (SINR) threshold increases, and thus
the average data rate keeps increasing in the lower and medium
regions of $\gamma_{\mathrm{th}}$. However, the successful access
probability decreases as $\gamma_{\mathrm{th}}$ increases, and thus
the increment of average data rate grows more slowly, or even
declines in the high $\gamma_{\mathrm{th}}$ region. Since the power
consumption is fixed, the trends of the energy efficiency curves
almost match their corresponding average data rate curves.

\subsection{Coordinated Scheduling}

\begin{figure}
\centering  \vspace*{0pt}
\includegraphics[scale=0.75]{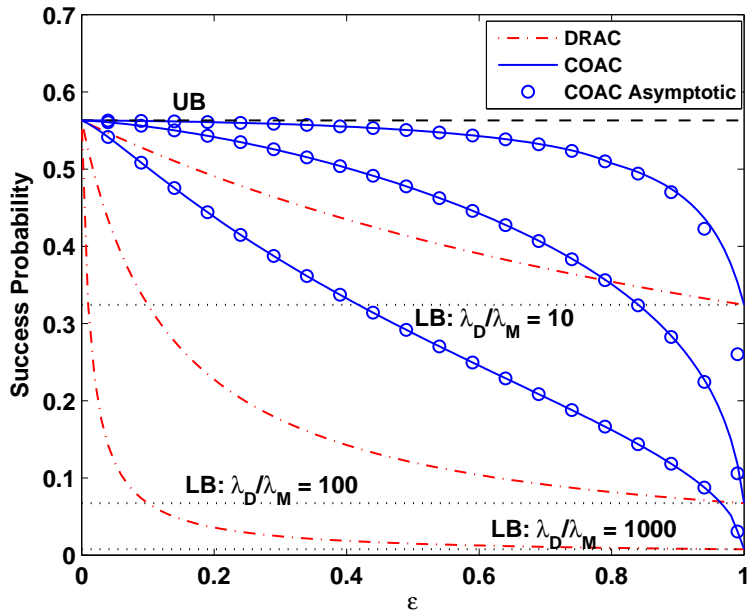}
\setlength{\belowcaptionskip}{-100pt} \caption{Cellular success
probability with respect to $\varepsilon$ under the DRAC and the
COAC schemes. The upper bound is shown with the dashed line, while
three lower bounds corresponding to ${\lambda _\mathrm{D}}/{\lambda
_\mathrm{M}}=$10, 100, and 1000 are provided with the dotted lines.}
\label{MUE_Success_Rate_with_Epsilon_Fig_6}\vspace*{-10pt}
\end{figure}

The coordinated scheduling is another emerging approach to mitigate
interference in the MAC for the D2D and relay mode, local
distributed coordination mode, and HPN mode. For example, to
mitigate interference between D2D F-UEs and F-UEs accessing F-APs,
the centralized opportunistic access control (COAC) is presented in
\cite{D2D_peng}, where each D2D F-UE opportunistically accesses the
sub-channels based on the centralized control by the HPN. The
performance comparison between COAC and the distributed random
access control (DRAC) are shown in
Fig.~\ref{MUE_Success_Rate_with_Epsilon_Fig_6} for various spectrum
occupation ratios of utilizing DRAC and COAC, i.e., $\varepsilon$.
Since DRAC and COAC have the same effect on the D2D success
probability, the performance is evaluated in terms of the cellular
success probability, where sparse, medium, and dense D2D densities
(denoted as ${\lambda _{\mathrm{D}}}/{\lambda _{\mathrm{M}}} = 10,
100, 1000$ with fixed ${\lambda _{\mathrm{M}}}$) are considered. In
addition, the asymptotic result is also plotted to show the
tightness of the results. For comparison, the case of $\varepsilon =
0$ and $\varepsilon = 1$ are presented as the upper and lower
bounds, respectively. It can be seen that the cellular success
probabilities decrease with the increase of $\varepsilon$, and the
extreme points exactly match with the upper and lower bounds that
correspond to $\varepsilon = 0$ and $\varepsilon = 1$, respectively.
Overall, COAC provides significant performance gains over DRAC by
exploiting the benefits of centralized management and opportunistic
accessing.

The optimal coordinated scheduling for F-RANs considering the
cross-layer optimization of multiple objective for the delay-aware
circumstance is often complex because there is diverse interference
among F-UEs with different transmission modes. To achieve the
optimal solution taking the queuing delay into account, the
equivalent rate constraint, Lyapunov optimization, and Markov
decision process (MDP) are three common tools. The equivalent rate
constraint approach converts the average delay constraints into
equivalent average rate constraints using queuing theory or large
deviation theory\cite{opt1}. The Lyapunov optimization approach
converts the average delay constraints into minimizing the Lyapunov
drift-plus-utility function \cite{opt2}. The Markov decision process
approach is a systematic approach to solving the derived Bellman
equation in a stochastic learning or differential equation manner.
In \cite{opt3}, to minimize the queuing delay under the average
power and fronthaul consumption constraints in the global C-RAN
mode, the queue-aware rate and power allocation problem is
formulated as an infinite horizon average cost constrained partially
observed Markov process decision, which takes both the urgent queue
state information and the imperfect channel state information at
transmitter (CSIT) into account. A stochastic gradient algorithm is
proposed to allocate power and transmission rate dynamically with
low computing complexity and high robustness against the variations
and uncertainties caused by unpredictable random traffic arrivals
and imperfect CSIT.

\section{Challenging Work and Open Issues}

Although F-RAN is a promising technology to copy with disadvantages
in C-RANs and H-CRANs, there are still some challenges and open
issues that remain to be discussed in the future, including the edge
caching, software-defined networking (SDN) and network function
virtualization (NFV) technologies.

\subsection{Edge Caching}

Caching aims to achieve a trade-off between the transmission rate
and the storage. If the transmission rate is high, the requirement
of storage is low. In F-RANs, though the scale of content acquired
by service providers is growing significantly, it is unnecessary to
cache all content in the cloud server, which increases the
end-to-end delay. Some local traffic are stored in edge devices can
significantly decrease burden on the constrained fronthaul, and
improve performances of CRSP and CRRM. With caching and computation
capabilities at edge devices, the edge caching is applied to relax
the traffic burden at the cloud server and provide the fast content
access and retrieval at F-UEs \cite{I:edgecloud}. Therefore, edge
caching is a key component to improve performances of F-RANs. The
key benefits of edge caching in F-RANs include: 1) the alleviation
of burden on the fronthaul, backhaul and even backbone; 2) the
reduction of content delivery latency; and 3) the flexibly
implementation of object-oriented or content-aware techniques to
improve F-RAN performances and user experiences.

Compared with the traditional centralized caching mechanism, the
caching space at each F-AP and F-UE is practically small, and the
edge caching often has a low-to-moderate hit ratio, which is the
probability that a requested content can be found in the cache.
Hence, intelligent caching resource allocation strategies and
cooperative caching policies among edge devices is mandatory.
Meanwhile, to exploit the edge caching benefits, some key factors
should be considered and jointly optimized, such as the cache load,
cache hit ratio, number of content requests, cost of caching
hardware, and cost of radio resource usage. Meanwhile, the caching
policies, deciding what to cache and when to release caches in
different edge devices, are crucial for improving the overall
caching performance. The traditional caching policies, such as
first-in first-out, least recently used, least frequently used,
should be evolved to appropriately improve the cache hit ratio in
F-RANs.

\subsection{Software-Defined Networking}

Inherited from H-CRANs, F-RAN decouples the control and user planes,
and the functions of F-APs can be re-configured by the software
stack with proprietary languages. Intuitively, since SDN decouples
the control plane from the data plane via controllers and allows
software to be designed independently from the hardware \cite{SDN},
F-RANs as the RAN of 5G systems can be seamlessly converged with SDN
as the core network. With the CRSP and CRRM procedures incorporated
into the edge devices in F-RANs, the SDN controlling, which is
originally designed for wired core networks, can be extended to the
physical layer in addition to the network layer, and more flexible
and efficient network control can be achieved. For the highly
flexible interfaces among different edge devices, SDN can be
recognized as a basic enabler to achieve the flexibility and
reconfigurability of F-RANs.

However, the data forwarding flow in SDN is mainly at the internet
protocol (IP) layer, and how to combine the functions of MAC and
physical layer for edge devices in F-RANs is still not
straightforward. Meanwhile, SDN is based on the centralized manner,
while the F-RAN has a high emphasis on the distributed manner for
edge devices. Therefore, it is challenging to achieve the SDN ideas
in real F-RANs. SDN for F-RANs needs to define slices which require
to isolate the CRSP and CRRM in edge devices so as to provide
non-interfering networks to different coordinators. The status and
locations of edge devices should be reported to SDN timely, based on
which the SDN controllers can take decisions efficiently, which is
also challenging because it will increase burden on fronthaul and
decrease advantages of F-RANs. These aforementioned challenges are
no-trivial and should be copied with for the successful rollout of
F-RANs with SDN.

\subsection{Network Function Virtualization}

NFV is the concept of transferring the network functions from
dedicated hardware appliances to software-based applications, which
aims to revolutionize the telecommunication industry by decoupling
network functions from the underlying proprietary hardware.
Recently, academic researchers and network engineers are exploiting
virtual environments to simplify and enhance NFV in order to find
its way smoothly into the telecommunications industry \cite{NFV}.

Based on the SDN for F-RANs, the programmable connectivity between
virtual network functions (VNFs) is provided and can be managed by
the orchestrator of VNFs which will mimic the role of the SDN
controller. Furthermore, NFV can virtualize the SDN controller to
run on the cloud server, which could be migrated to fit locations
according to the network needs. However, how to virtualize the SDN
controller in F-RANs is still indistinct due to the distribution
characteristic in edge devices. The security, computing performance,
VNF interconnection, portability, compatible operation and
management with legacy RANs specified for F-RANs are main challenges
and should be exploited in the future.

\section{Conclusion}

In this article, we have introduced a fog computing based radio
access network architecture for 5G systems, which incorporates fog
computing into H-CRANs. Compared with the traditional centralized
cloud computing based C-RANs/H-CRANs, cooperative radio signal
processing and cooperative radio resource management procedures in
F-RANs are adaptively implemented at the edge devices and are closer
to the end users. With the goal of understanding further intricacies
of key techniques, we have presented transmission mode selection and
interference suppression. Within these two key techniques, we have
summarized the diverse problems and the corresponding solutions that
have been proposed. Nevertheless, given the relative infancy of the
field, there are still quite a number of outstanding problems that
need further investigation. Notably, it is concluded that greater
attention should be focused on transforming the F-RAN paradigm into
edge caching, SDN and NFV.

\begin{IEEEbiography}{Mugen Peng}
(M'05--SM'11) received the PhD degree in Communication and
Information System from the Beijing University of Posts \&
Telecommunications (BUPT), China, in 2005. Now he is a full professor
with the school of information and communication engineering in
BUPT. His main research areas include cooperative communication,
heterogeneous network, and cloud communication. He has
authored/coauthored over 50 refereed IEEE journal papers and over
200 conference proceeding papers. He received the 2014 IEEE ComSoc
AP Outstanding Young Researcher Award, and the Best Paper Award in
IEEE WCNC 2015, GameNets 2014, IEEE CIT 2014, ICCTA 2011, IC-BNMT 2010, and IET CCWMC
2009.
\end{IEEEbiography}

\begin{IEEEbiography}{Yan Shi} is currently a PhD candidate with the Key Laboratory of
Universal Wireless Communication (Ministry of Education) at Beijing University of Posts \& Telecommunications (BUPT), China.
His research interest focuses on the performance analysis and optimization of cloud and fog computing based radio access networks.
\end{IEEEbiography}

\begin{IEEEbiography}{Kecheng Zhang} is currently a PhD candidate with the Key Laboratory of
Universal Wireless Communication (Ministry of Education) at Beijing University of Posts \& Telecommunications (BUPT), China.
His research interest focuses on the radio resource allocation of fog computing based radio access networks.
\end{IEEEbiography}

\begin{IEEEbiography} {Chonggang Wang}
(SM'09) received his Ph.D. degree from BUPT in 2002. He is a member
technical staff with InterDigital Communications focusing on
Internet of Things (IoT) R \& D activities, including technology
development and standardization. His current research interests
include IoT, mobile communication and computing, and big data
management and analytic. He is the founding Editor-in-Chief of
\emph{IEEE Internet of Things Journal} and on the editorial board of
several journals, including IEEE Access.
\end{IEEEbiography}

\end{document}